\documentclass[english,aip, manuscript]{revtex4-1}

\pdfoutput=1
\usepackage[T1]{fontenc}
\usepackage[latin9]{inputenc}
\usepackage{color}
\usepackage{babel}
\usepackage{textcomp}
\usepackage{amsmath}
\usepackage{amssymb}
\usepackage{graphicx}
\usepackage{esint}
\usepackage{subscript}
\usepackage[unicode=true,pdfusetitle,
 bookmarks=true,bookmarksnumbered=false,bookmarksopen=false,
 breaklinks=false,pdfborder={0 0 1},backref=false,colorlinks=true]
 {hyperref}
\usepackage{breakurl}

\makeatletter

\providecommand{\tabularnewline}{\\}

 \@ifundefined{textcolor}{}
 {%
   \definecolor{BLACK}{gray}{0}
   \definecolor{WHITE}{gray}{1}
   \definecolor{RED}{rgb}{1,0,0}
   \definecolor{GREEN}{rgb}{0,1,0}
   \definecolor{BLUE}{rgb}{0,0,1}
   \definecolor{CYAN}{cmyk}{1,0,0,0}
   \definecolor{MAGENTA}{cmyk}{0,1,0,0}
   \definecolor{YELLOW}{cmyk}{0,0,1,0}
 }

\def\part#1{\left(#1\right)}

\usepackage{amsthm}
\usepackage{graphics}
\usepackage{times}
\usepackage{bm}
\usepackage{hyperref}
\usepackage{color}

\RequirePackage[dvipsnames]{xcolor} 
\definecolor{RoyalBlue}{cmyk}{1, 0.80, 0, 0}

\hypersetup{
colorlinks=true,
breaklinks=true, 
hyperfootnotes=true,
urlcolor=RoyalBlue, 
citecolor=RoyalBlue, 
linkcolor=RoyalBlue, 
}

\@ifundefined{showcaptionsetup}{}{%
 \PassOptionsToPackage{caption=false}{subfig}}
\usepackage{subfig}
\makeatother

\begin{document}

\title{Graphics Processing Units Accelerated Semiclassical Initial Value
Representation Molecular Dynamics}

\author{Dario Tamascelli and Francesco Saverio Dambrosio}

\affiliation{Dipartimento di Fisica, Università degli Studi di Milano, via Celoria
16, 20133 Milano, Italy}

\author{Riccardo Conte}

\affiliation{Department of Chemistry and Cherry L. Emerson Center for Scientific
Computation, Emory University, Atlanta, Georgia 30322, United States}

\author{Michele Ceotto}

\affiliation{Dipartimento di Chimica, Università degli Studi di Milano, via Golgi
19, 20133 Milano, Italy}

\email{michele.ceotto@unimi.it}

\selectlanguage{english}%
\begin{abstract}
This paper presents a Graphics Processing Units (GPUs) implementation
of the Semiclassical Initial Value Representation (SC-IVR) propagator
for vibrational molecular spectroscopy calculations. The time-averaging
formulation of the SC-IVR for power spectrum calculations is employed.
Details about the GPU implementation of the semiclassical code are
provided. Four molecules with an increasing number of atoms are considered
and the GPU-calculated vibrational frequencies perfectly match the
benchmark values. 
The computational time scaling of two GPUs (NVIDIA Tesla C2075 and Kepler K20) respectively versus two CPUs (Intel Core i5 and Intel Xeon E5-2687W) and the critical issues related to the GPU implementation are discussed. The resulting reduction in computational time and power consumption is significant and semiclassical GPU calculations are shown to be environment friendly.
\end{abstract}
\maketitle

\section{Introduction\label{sec:Introduction}}

The exponentially increasing demand for advanced graphics solutions
for many software applications including entertainment, visual simulation,
computer-aided design and scientific visualization has boosted high-performance
graphics systems architectural innovation.\cite{NVIDIA} Nowadays,
Graphics Processing Units (GPUs) are ubiquitous, affordable and designed
to exploit the tremendous amount of data parallelism of graphics algorithms. 

In recent years, GPUs have evolved into fully programmable devices
and they are now ideal resources for accelerating several scientific
applications. GPUs are designed with a philosophy which is very different
from CPUs. On one hand, CPUs are more flexible than GPUs and able
to provide a fast response to a single task instruction. On the other
hand, GPUs are best performing for highly parallelized processes.
CPUs provide caches and this hardware tool has been developed in a
way to better assist programmers. In particular, caches are transparent
to programmers and the recent bigger ones can capture most used data.
GPUs, instead, achieve high performances by means of hundreds of cores
which are fed by multiple independent parallel memory systems. A single
GPU is composed of groups of single-instruction multiple-data (SIMD)
processing units and each unit is made of multiple smaller processing
parts called threads. These are set to execute the same instructions
concurrently. The advantages of this type of architecture consist
in a reduced power consumption and an increased number of floating
point arithmetic units per unit area. In other words, a reduced amount
of space, power and cooling is necessary to operate. However, parallelization
efficiency depends critically on threads synchronization. In fact,
accidental or forced inter-thread synchronization can turn out to
be very costly, because it involves a kernel termination. Generation
of a new kernel implies overhead from the host. Another GPU drawback
is represented by its better efficiency for single precision arithmetics.
Unfortunately, single precision is not enough for most scientific
calculations. In general, then, there may not be a single stable GPU
programming model and CPU codes need usually to be extensively changed
in order to fit the GPU hardware.

GPUs are becoming more and more popular among the scientific community
mainly thanks to the release of NVIDIA's Compute Unified Device Architecture
(CUDA) toolkit.\cite{CUDA} This is a programming model based on a
user-friendly decomposition of the code into grids and threads which
significantly simplifies the code development. It allows to exploit
all of the key hardware capabilities, such as scatter/gather and thread
synchronizations.

Applications of GPU programming in theoretical chemistry include implementations
for classical molecular dynamics (MD),\cite{Schulten_GPUMD_07,Schulten_2,Schulten_conference,Elber2}
quantum chemistry \cite{Martinez_2008,Yasuda,Alan_JPCA_08,Genovese,Alan_QuantumChemistryGPU_10,Tomono,Genovese2,Alan_Rubio_12,Dronskowski,Hammond,QE_GPU,Stotzka,Maia,Hacene,Esler,Hakala,Jia,Hutter},
protein folding \cite{Pande_GPUfolding_13}, quantum dynamics\cite{Zerbetto_QMGPU_12,Han_scattering_13,Lagana1,Lagana2,Lagana3,Murgante}
and quantum mechanics / molecular mechanics (QM/MM) \cite{Maezono_QMMMGPU_11}
simulations. For instance, classical MD can be sped up by using GPUs
for the calculation of long-range electrostatics and non-bonded forces.\cite{Schulten_GPUMD_07,Schulten_2}
The direct Coulomb summation algorithm accesses the shared memory
area only at the very beginning and the very end of the processing
for each thread block, so MD takes full advantage of the GPUs architecture
by eliminating any use of thread synchronizations. For instance, the
popular Not (just) Another Molecular Dynamics (NAMD) program \cite{Schulten_2,Schulten_conference}
is accelerated several times by directing the electrostatics and implicit
solvent model calculations to GPUs while the remaining tasks are handled
by CPUs. Significant progress by Friedrichs et al. has determined
a MD speed up of about 500 times over an 8-core CPU by using the OpenMM
library.\cite{Friedrichs} More difficult has been the adoption of
GPUs for quantum chemistry. The first full electronic-structure implementation
on GPUs was Ufimtsev and Martinez's TeraChem.\cite{Martinez_2008}
Currently, there are several electronic structure codes that have
to some extent implemented GPU accelerations.\cite{Yasuda,Alan_QuantumChemistryGPU_10,QE_GPU}
For example, Aspuru's group successfully accelerated real-space DFT
calculations making this approach interesting and competitive.\cite{Alan_JCTC_13_realspaceDFT}
A simple GPU implementation of the Cublas SGEMM subroutine in quantum
chemistry has been shown to be about 17 times faster than the parent
DGEMM subroutine on CPU.\cite{Alan_QuantumChemistryGPU_10} Recently,
CPU/GPU-implemented time-independent quantum scattering calculations
featured a 7-time acceleration by employing 3 GPU and 3 CPU cores. 

In a time-dependent quantum propagation, instead, almost all of the
computational resources are spent for the time propagation of the
wavepacket. Only the initial wavepacket is calculated on the CPU.
Data are copied from the CPU memory to the GPU one for wavepacket
propagation. Furthermore, it has been shown that quantum time-dependent
approaches can be boosted up to two orders of magnitude by taking
advantage of the matrix-matrix multiplication for the time-evolution
that maps well to GPU architectures.\cite{Zerbetto_QMGPU_12,Han_scattering_13}
Lagana's group demonstrated that quantum reactive scattering for reactive
probabilities calculations can be accelerated as much as 20 times.\cite{Lagana1,Lagana2,Lagana3,Lagana_Pacifici_13}

The main goal of this paper is to speed up our semiclassical dynamics
CPU code by exploiting the GPU hardware. We show how and when it is
convenient to employ GPU devices to perform semiclassical simulations.
The GPU approach is also demonstrated to require a largely reduced
amount of power supply. Unfortunately, GPU accelerated programming
experiences gained for quantum propagation matrix-matrix multiplications
or for the classical Coulombic MD force field are not helpful in the
case of semiclassical simulations, due to the need to calculate concurrently
quantum delocalization and classical localization. Given the mixed
classical and quantum nature of the semiclassical propagator, a general
purpose (GP) GPU approach is taken. With this approach host codes
run on CPUs and kernel codes on GPUs. GPGPU programming is principally
aimed at minimizing data transfer between the host and the kernel,
since this communication is made via bus with relatively low speed. 

The paper is organized as follows. Next Section recalls the semiclassical
initial value representation quantum propagator and the following
one describes the GPGPU programming approach adopted here. The Section
compares the performances of CPU and GPGPU codes and discusses them.
The last Section reports our conclusions.

\section{Semiclassical initial value representation of the quantum propagator}

The semiclassical propagator can be derived from the Feynman Path
Integral formulation of the quantum evolution operator \cite{Feynman_Hibbs}
from point \textbf{$\mathbf{q}$} to $\mathbf{q}^{\prime}$ 
\begin{equation}
\mbox{\ensuremath{\left\langle \mathbf{q}^{\prime}\left|e^{-i\hat{H}t/\hbar}\right|\mathbf{q}\right\rangle }}=\left(\frac{m}{2\pi i\hbar t}\right)^{1/2}\int\mathcal{D}\left[\mathbf{q}\left(t\right)\right]e^{iS_{t}\left(\mathbf{q},\mathbf{q}^{\prime}\right)/\hbar}\label{eq:FeynmanPaths}
\end{equation}
where $S_{t}\left(\mathbf{q},\mathbf{q}^{\prime}\right)$ is the path
action for time $t$ and $\mathcal{D}\left[\mathbf{q}\left(t\right)\right]$
indicates the differential over all paths. Stationary phase approximation
of Eq. (\ref{eq:FeynmanPaths}) (see, for instance, Ref. \citenum{Berry_Mount, Tannor_book})
yields the semiclassical van Vleck-Gutzwiller propagator\cite{vanVleck_PNAS,Gutzwiller_SCpropagator}
\begin{equation}
\left\langle \mathbf{q}^{\prime}\left|e^{-i\hat{H}t/\hbar}\right|\mathbf{q}\right\rangle \approx\sum_{\mbox{roots}}\left[\frac{1}{\left(2\pi i\hbar\right)^{F}}\left|-\frac{\partial^{2}S}{\partial\mathbf{q}^{\prime}\partial\mathbf{q}}\right|\right]^{1/2}e^{iS_{t}\left(\mathbf{q},\mathbf{q}^{\prime}\right)/\hbar-i\nu\pi/2}\label{eq:vanVleck}
\end{equation}
where the sum is over all classical trajectories going from $\mathbf{q}$
to $\mathbf{q}^{\prime}$ in an amount of time $t$, $F$ is the number
of degrees of freedom, and $\nu$ is the Maslov or Morse index, i.e.
the number of points along the trajectory where the determinant in Eq. (\ref{eq:vanVleck}) diverges.\cite{Morse,Maslow}
To apply Eq. (\ref{eq:vanVleck}) as written, one needs to solve a
nonlinear boundary value problem. The classical trajectory evolved
from the initial phase space point $\left(\mathbf{p}\left(0\right),\mathbf{q}\left(0\right)\right)$ is
such that $\mathbf{q}_{t}\left(\mathbf{p}\left(0\right),\mathbf{q}\left(0\right)\right)=\mathbf{q}^{\prime}$.
In general, there will be multiple roots to this equation and the
summation of Eq. (\ref{eq:vanVleck}) is over all such roots. Finding
these roots is a formidable task that has hindered use and diffusion
of semiclassical dynamics. The issue was overcome by Miller's Semiclassical
Initial Value Representation (SC-IVR), whereby the boundary condition
summation is replaced by an initial phase space integration amenable
to Monte Carlo implementation.\cite{Miller_IVR,Grossmann_avd,Kay_IVR,Pollak,ContePollak10,ContePollak12,Jiri}
By representing the van Vleck-Gutzwiller propagator by direct product
of one-dimensional $\gamma_{i}$-width coherent states \cite{Kay_IVR,Heller_frozengaussian,HermanKlukcoherstates,SunMiller_HCldimer,Miller_JPC_featurearticle,Miller_PNAS}
defined by 
\begin{align}
\left\langle \mathbf{q}|\mathbf{p}\left(t\right),\mathbf{q}\left(t\right)\right\rangle  & =\prod_{i}\left(\mathbf{\gamma}_{i}/\pi\right)^{F/4}\nonumber \\
\times & \mbox{exp}\left[-\frac{\gamma_{i}}{2}\left(q_{i}-q_{i}\left(t\right)\right)^{2}+\frac{i}{\hbar}p_{i}\left(t\right)\left(q_{i}-q_{i}\left(t\right)\right)\right]\label{eq:coherent_state}
\end{align}
and using Miller's IVR trick, the semiclassical propagator becomes

\textbf{
\begin{eqnarray}
e^{-i\hat{H}t/\hbar}= & \frac{1}{\left(2\pi\hbar\right)^{F}}\int d\mathbf{p}\left(0\right)\int d\mathbf{q}\left(0\right)\: C_{t}\left(\mathbf{p}\left(0\right),\mathbf{q}\left(0\right)\right)\nonumber \\
 & e^{iS_{t}\left(\mathbf{p}\left(0\right),\mathbf{q}\left(0\right)\right)/\hbar}\left|\mathbf{p}\left(t\right),\mathbf{q}\left(t\right)\left\rangle \right\langle \mathbf{p}\left(0\right),\mathbf{q}\left(0\right)\right|.\label{eq:SCTime_evolution}
\end{eqnarray}
}$\left(\mathbf{p}\left(t\right),\mathbf{q}\left(t\right)\right)$
represent the set of classically-evolved phase space coordinates and
$C_{t}$ is a pre-exponential factor. In the Herman-Kluk frozen Gaussian
version of SC-IVR, the pre-exponential factor is written as\cite{Kay_IVR,Heller_frozengaussian,HermanKlukcoherstates}
\begin{eqnarray}
C_{t}\left(\mathbf{p}\left(0\right),\mathbf{q}\left(0\right)\right) & =\label{eq:prefactor}\\
\sqrt{\frac{1}{2}\left|\frac{\partial\mathbf{q}\left(t\right)}{\partial\mathbf{q}\left(0\right)}+\frac{\partial\mathbf{p}\left(t\right)}{\partial\mathbf{p}\left(0\right)}-i\hbar\Gamma\frac{\partial\mathbf{q}\left(t\right)}{\partial\mathbf{p}\left(0\right)}+\frac{i}{\Gamma\hbar}\frac{\partial\mathbf{p}\left(t\right)}{\partial\mathbf{q}\left(0\right)}\right|}\nonumber 
\end{eqnarray}
where $\Gamma=\mbox{diag}\left(\gamma_{1},...,\gamma_{F}\right)$
is the coherent state matrix which defines the Gaussian width of the
coherent state. The calculation of C\textsubscript{t} is conveniently
performed from blocks of size $F\times F$ by introducing a $2F\times2F$
symplectic (monodromy or stability) matrix $\mathbf{M}\left(t\right)\equiv\partial\left(\left(\mathbf{p}_{t},\mathbf{q}_{t}\right)/\partial\left(\mathbf{p}_{0},\mathbf{q}_{0}\right)\right)$.
The accuracy of time-evolved classical trajectories is monitored by
calculating the deviation of the determinant of the positive-definite
matrix $\mathbf{M}^{T}\mathbf{M}$ from unity.\cite{Miller340_GeneralizedFilinov_01}
In this work, a trajectory is discarded when its deviation is greater
than $10^{-6}$. For semiclassical dynamics of bound systems, a reasonable
choice for the $\gamma_{i}$ width parameters is provided by the harmonic
oscillator approximation to the wave function at the global minimum.

In this paper, we employ the SC-IVR propagator to calculate the spectral
density 
\begin{equation}
I\left(E\right)\equiv\left\langle \chi\left|\delta\left(\hat{H}-E\right)\right|\chi\right\rangle =\sum_{n}\left|\left\langle \chi|\psi_{n}\right\rangle \right|^{2}\delta\left(E-E_{n}\right),\label{eq:spectrum}
\end{equation}
where $\left|\chi\right\rangle $ is some reference state, $\left\{ \left|\psi_{n}\right\rangle \right\} $
are the exact eigenfunctions and $\left\{ E_{n}\right\} $ the corresponding
eigenvalues of the Hamiltonian $\hat{H}$. A more practical dynamical
representation of Eq. (\ref{eq:spectrum}) is given by the following
time-dependent representation\cite{Heller_review_autocorrel}, 
\begin{equation}
I\left(E\right)=\frac{1}{2\pi\hbar}\int_{-\infty}^{+\infty}\left\langle \chi\left|e^{-i\hat{H}t/\hbar}\right|\chi\right\rangle e^{iEt/\hbar}dt=\frac{\mbox{Re}}{\pi\hbar}\int_{0}^{+\infty}\left\langle \chi\left|e^{-i\hat{H}t/\hbar}\right|\chi\right\rangle e^{iEt/\hbar}dt\label{eq:I(E)_timedependent}
\end{equation}
which is obtained by replacing the Dirac delta function in Eq. (\ref{eq:spectrum})
by its Fourier representation. According to Eq. (\ref{eq:I(E)_timedependent})
and Eq. (\ref{eq:SCTime_evolution}), the SC-IVR spectral density
representation becomes\cite{Miller312_FaradayDisc_98}

\begin{eqnarray}
I\left(E\right) & = & \frac{1}{2\pi\hbar}\int_{-\infty}^{+\infty}e^{iEt/\hbar}\frac{1}{\left(2\pi\hbar\right)^{F}}\int d\mathbf{p}\left(0\right)\int d\mathbf{q}\left(0\right)\: C_{t}\left(\mathbf{p}\left(0\right),\mathbf{q}\left(0\right)\right)\nonumber \\
 & \times & e^{iS_{t}\left(\mathbf{p}\left(0\right),\mathbf{q}\left(0\right)\right)/\hbar}\left\langle \chi|\mathbf{p}\left(t\right),\mathbf{q}\left(t\right)\right\rangle \left\langle \mathbf{p}\left(0\right),\mathbf{q}\left(0\right)|\chi\right\rangle dt\label{eq:I(E)_SCIVR}
\end{eqnarray}
where the reference state $\left|\chi\right\rangle =\left|\mathbf{p}_{eq},\mathbf{q}_{eq}\right\rangle $
is represented in phase space coordinates. The Monte Carlo phase space
integration is made easier to treat by introducing a time averaging
filter at the cost of a longer simulation time. This implementation
was introduced by Kaledin and Miller\cite{Alex} resulting in the
following time-averaging (TA) SC-IVR formulation for the spectral
density

\begin{eqnarray}
I\left(E\right) & = & \frac{1}{\left(2\pi\hbar\right)^{F}}\int d\mathbf{p}\left(0\right)\int d\mathbf{q}\left(0\right)\frac{\mbox{Re}}{\pi\hbar T}\int_{0}^{T}dt_{1}\int_{t_{1}}^{T}dt_{2}\: C_{t_{2}}\left(\mathbf{p}\left(t_{1}\right),\mathbf{q}\left(t_{1}\right)\right)\nonumber \\
 & \times & \left\langle \chi\left|\right.\mathbf{p}\left(t_{2}\right),\mathbf{q}\left(t_{2}\right)\right\rangle e^{i\left(S_{t_{2}}\left(\mathbf{p}\left(0\right),\mathbf{q}\left(0\right)\right)+Et_{2}\right)/\hbar}\left[\left\langle \chi\left|\right.\mathbf{p}\left(t_{1}\right),\mathbf{q}\left(t_{1}\right)\right\rangle e^{i\left(S_{t_{1}}\left(\mathbf{p}\left(0\right),\mathbf{q}\left(0\right)\right)+Et_{1}\right)/\hbar}\right]^{*}.\label{eq:TA_spectrdens}
\end{eqnarray}
Eq. (\ref{eq:TA_spectrdens}) presents two time variables. The integration
over $t_{2}$ is taking care of the Fourier transform of Eq. (\ref{eq:I(E)_timedependent})
(limited to the simulation time T), while the one over $t_{1}$ does
the filtering job. The positions $\left(\mathbf{p}\left(t_{1}\right),\mathbf{q}\left(t_{1}\right)\right)$
and $\left(\mathbf{p}\left(t_{2}\right),\mathbf{q}\left(t_{2}\right)\right)$
are referred to the same trajectories but at different times. By adopting
a reasonable approximation for the pre-exponential factor, $C_{t_{2}}\left(\mathbf{p}\left(t_{1}\right),\mathbf{q}\left(t_{1}\right)\right)=\mbox{Exp}\left[i\left(\phi\left(t_{2}\right)-\phi\left(t_{1}\right)\right)/\hbar\right],$\cite{Alex}
where $\phi\left(t\right)=\mbox{phase}\left[C_{t}\left(\mathbf{p}\left(0\right),\mathbf{q}\left(0\right)\right)\right]$,
the double-time integration of Eq. (\ref{eq:TA_spectrdens}) is reduced
to a single one and the spectral density becomes 
\begin{eqnarray}
I\left(E\right) & = & \frac{1}{\left(2\pi\hbar\right)^{F}}\frac{1}{2\pi\hbar T}\int d\mathbf{p}\left(0\right)\int d\mathbf{q}\left(0\right)\nonumber \\
 & \times & \left|\int_{0}^{T}dt\left\langle \chi|\mathbf{p}\left(t\right),\mathbf{q}\left(t\right)\right\rangle \right.\label{eq:sep_approx}\\
 & \times & \left.e^{i\left(S_{t}\left(\mathbf{p}\left(0\right),\mathbf{q}\left(0\right)\right)+Et+\phi_{t}\left(\mathbf{p}\left(0\right),\mathbf{q}\left(0\right)\right)\right)/\hbar}\right|^{2}.\nonumber 
\end{eqnarray}
Eq. (\ref{eq:sep_approx}) offers the advantage that the integrand
is now positive-definite and the integration is less computationally
demanding. Several applications\cite{Alex,Ceotto_1traj,Ceotto_MCSCIVR,Ceotto_david,Ceotto_eigenfunctions,Ceotto_cursofdimensionality_11,Ceotto_acceleratedSCIVR,Ceotto_NH3}
have demonstrated that this approximation is quite accurate.

\section{GPU Implementation of the SC-IVR Spectral Density}

\subsection{Monte-Carlo SC-IVR algorithm}

\begin{figure}
\begin{centering}
\includegraphics[scale=0.5]{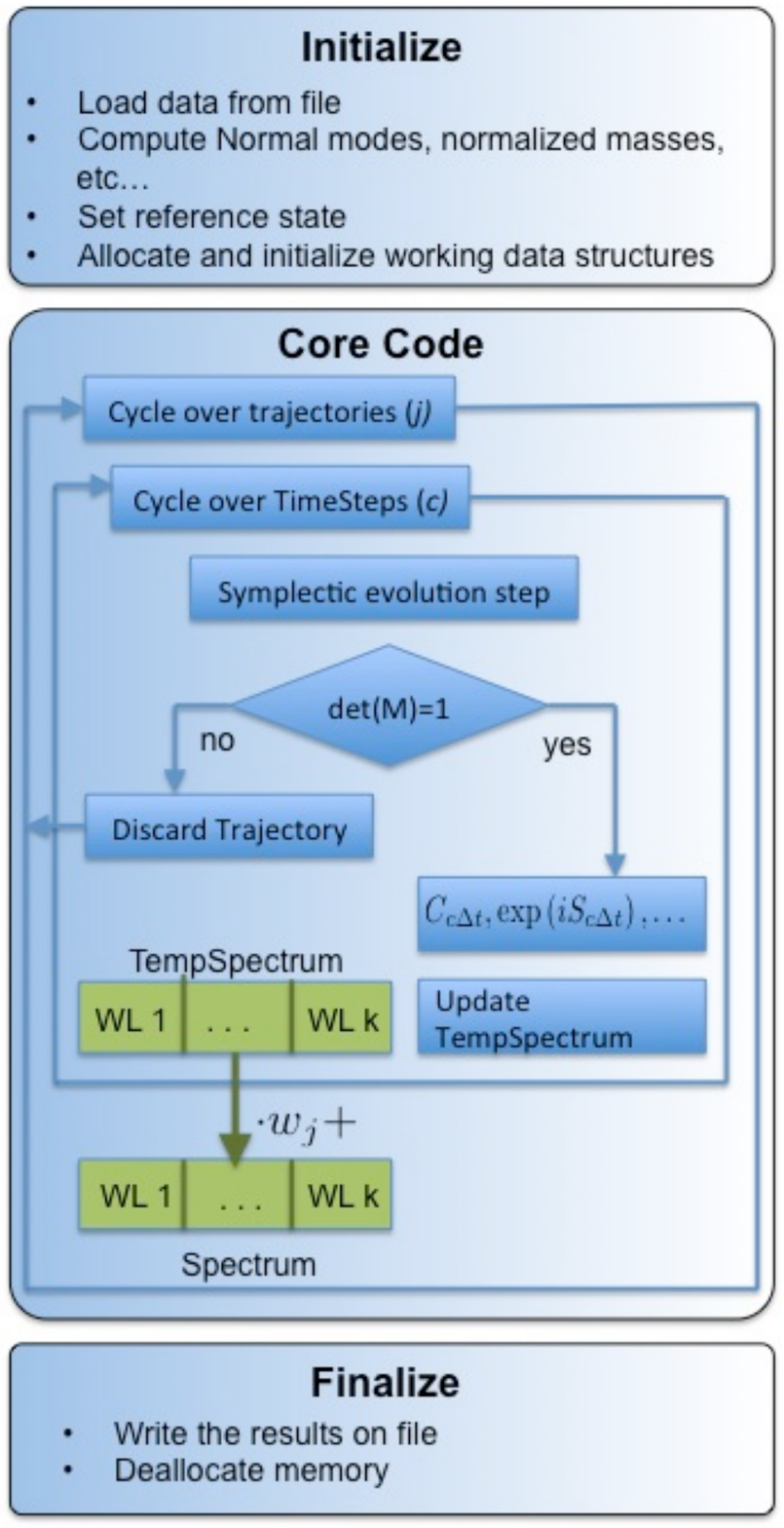}
\par\end{centering}

\caption{\label{fig:structureCPP}The structure of the sequential code.}
\end{figure}

To point out the degree of parallelism available in the SC-IVR procedure,
we describe the main steps that lead to the computation of the semiclassical
power spectrum. \\
The spectrum is conveniently represented as a $k$-dimensional vector
$\left(E_{1}=E_{min},E_{2},...,E_{k}=E_{max}\right)$ of equally spaced
points in the range $[E_{min},E_{max}]$. To evaluate each element
of the discretized spectrum we need to calculate $I(E_{i}),\, i=1,2,\ldots,k$
from Eq. (\ref{eq:sep_approx}). To this end we use a Monte Carlo
(MC) method. The phase-space integral of Eq. (\ref{eq:sep_approx})
is approximated by means of the following MC sum of $n_{traj}$ classical
trajectories

\begin{align}
I(E_{i})=\frac{1}{\left(2\pi\right)^{F+1}}\frac{1}{n_{good}T}\sum_{j=1}^{n_{traj}}w_{j}\left|\sum_{c=0}^{n_{steps}}\left\langle \chi\left|\mathbf{p}_{j}(c\Delta t),\mathbf{q}_{j}(c\Delta t)\right.\right\rangle e^{i\left(S_{c\Delta t}\left(\mathbf{p}_{j}\left(0\right),\mathbf{q}_{j}\left(0\right)\right)+E_{i}c\Delta t+\phi_{c\Delta t}\left(\mathbf{p}_{j}\left(0\right),\mathbf{q}_{j}\left(0\right)\right)\right)}\right|^{2}\label{eq:I(E)_discreto}
\end{align}
where atomic units have been employed, and $n_{good}$ is the actual
number of trajectories ($n_{good}\leq n_{traj}$) over which the sum
is averaged, as discussed at the end of next paragraph. The MC phase
space sampling is performed according to the Husimi distribution which
determines the weight $w_{j}$ of each trajectory.\cite{Alex} The
classical trajectories $\left(\mathbf{p}_{j}(t),\mathbf{q}_{j}(t)\right)$
and the actions are then evolved, through $n_{steps}$ discrete time
steps of length $\Delta t$ from time $0$ to time $T$ by means of
a fourth order symplectic algorithm.\cite{Ceotto_HessianAccuracy}\\
The structure of the sequential (CPU) code is shown in Fig. \ref{fig:structureCPP}.
 First, all the relevant information about the molecule under investigation
are read from the configuration files. These are the masses and the
equilibrium positions of the atoms. Then, normal mode coordinates
are generated together with the conversion matrix from normal modes
to Cartesian coordinates. This matrix is necessary, since the simulations
are performed in normal mode coordinates, while the potential subroutines
are written in Cartesian or Internal coordinates. Once all the simulation
and molecule-configuration parameters have been loaded into the program,
the sequential generation of MC trajectories starts. In order to check
the stability of the symplectic evolution of each trajectory, the
determinant of the monodromy matrix $\left|M(t)\right|$ is evaluated.
As soon as it deviates from unity by an amount greater than $10^{-6}$,
the evolution of the trajectory is interrupted and its contribution
to the MC integration discarded. The intermediate results produced
by the trajectory are stored in a buffer \texttt{(TempSpectrum}).
The buffered results contribute to the computation of the spectrum\texttt{(}which
is done in the\texttt{ Spectrum} array) only if the trajectory has
completed its evolution over the whole $[0,T]$ time interval. We
use a counter $n_{good}$ to count the number of \textquotedbl{}good\textquotedbl{}
trajectories. When all the trajectories have been generated, the spectrum
is normalized over $n_{good}$ and copied into a file.

\subsection{GP-GPU Implementation}

\begin{figure}
\subfloat[\label{fig:structureCUDA}CUDA code structure. The $j$-th row $(TW(j,1),TW(j,2),\ldots,TW(j,k))$
of the \texttt{Traj-WaveLength} matrix contains the private copy of
the \texttt{TempSpectrum} array of the $j$-th trajectory. After the
threads of \texttt{KernelEvolution} have filled the matrix, \texttt{KernelSpectrum}
is launched. Each thread of \texttt{KernelSpectrum} performs the sum
of the elements over each column of \texttt{Traj-WaveLength.}]{\begin{centering}
\includegraphics[scale=0.45]{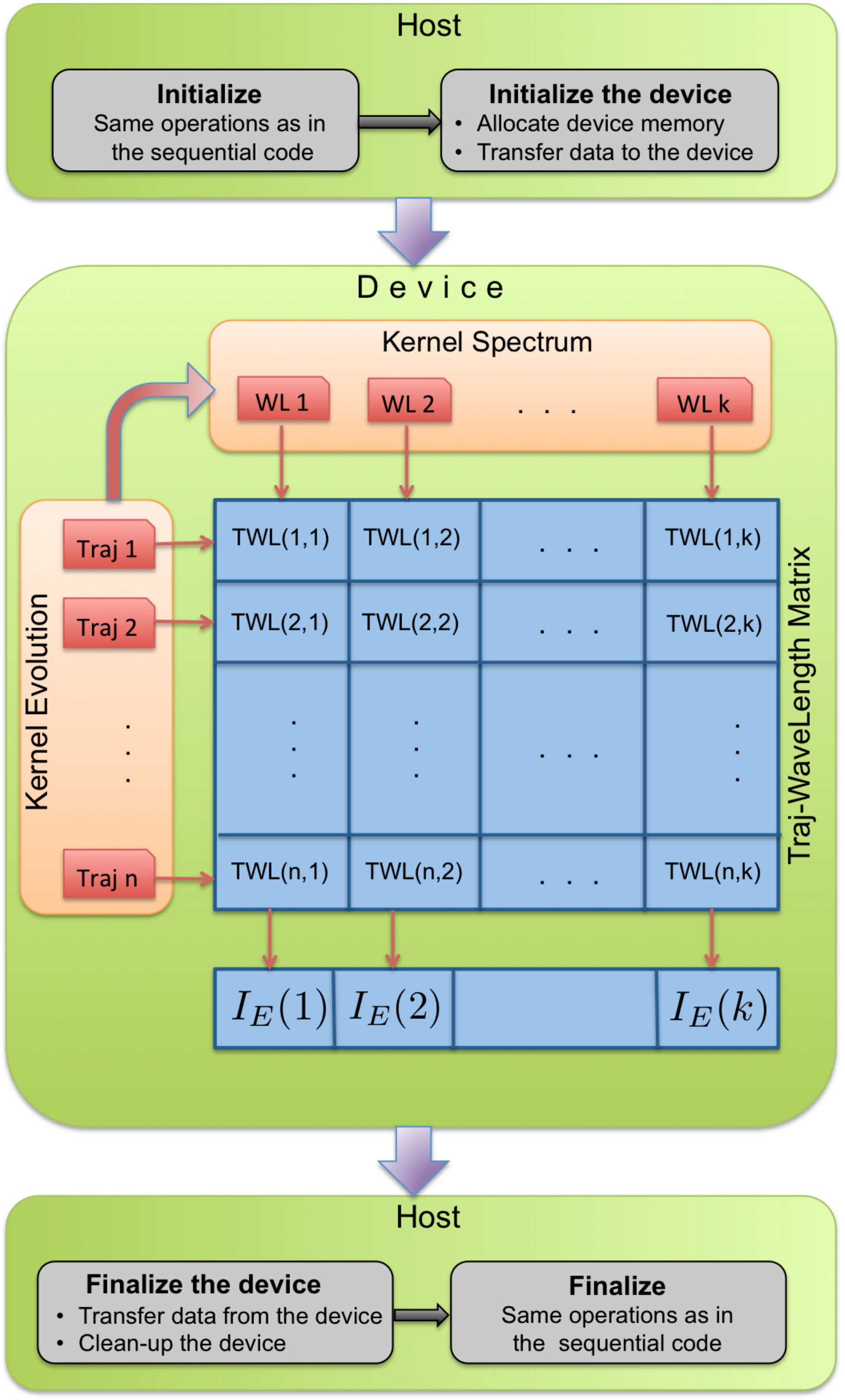}
\par\end{centering}

}\hfill{}\subfloat[\label{fig:figureMemoryCUDA}Use of the memory hierarchy while executing
the \textit{Evolution Kernel}.]{\begin{centering}
\includegraphics[width=0.45\columnwidth]{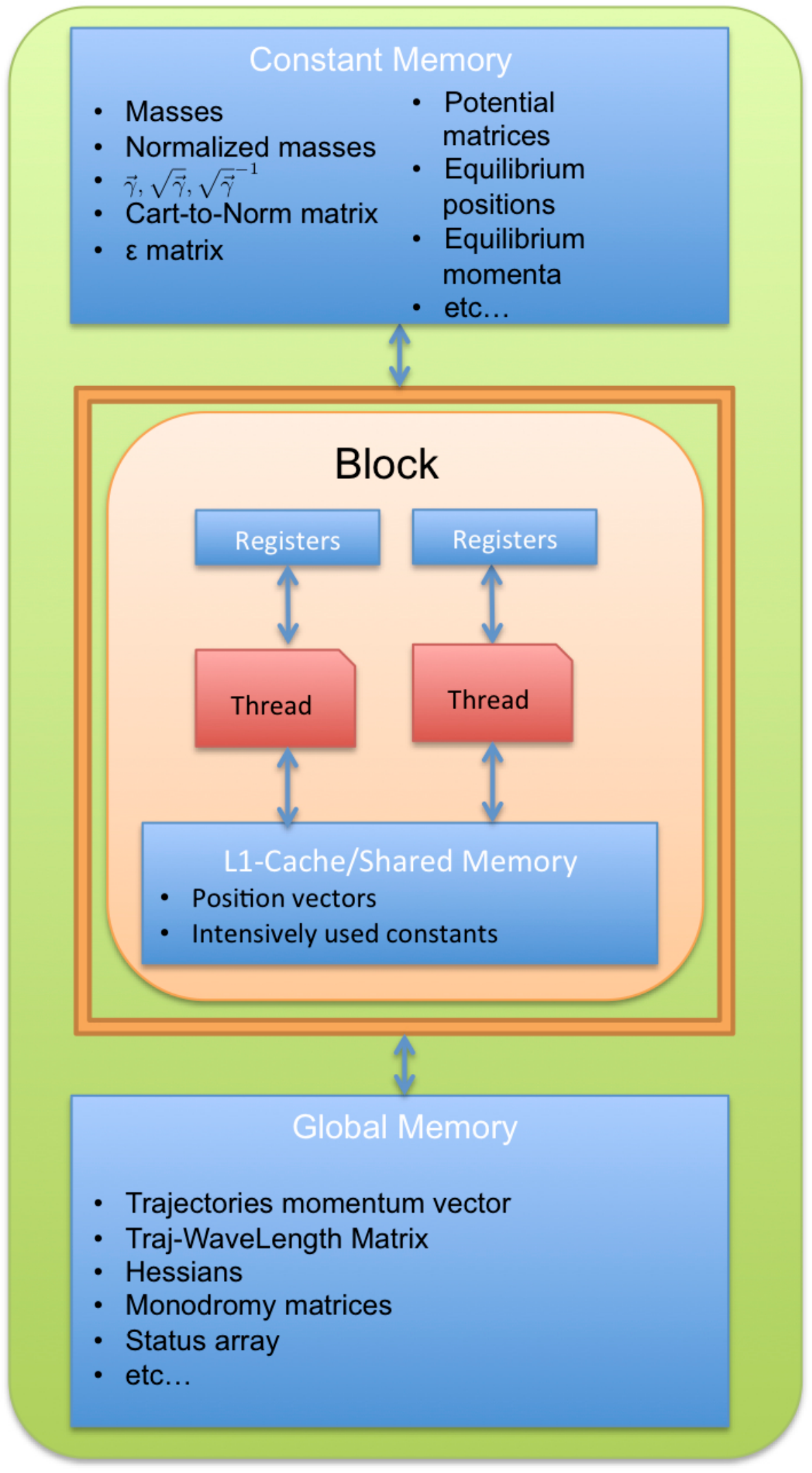}
\par\end{centering}

}

\caption{}
\end{figure}
 Since each trajectory evolves independently, the design of a parallel
version of the MC algorithm described in Eq. (\ref{eq:I(E)_discreto})
is rather straightforward. As the most direct implementation, a $n_{traj}$
simulation can be performed by means of $n_{traj}$ independent computational
units, each one working on a private memory space. Once all the trajectories
have been run, the results can be summed up to obtain the final spectrum.
Here, we describe the implementation of the MC SC-IVR algorithm for
$\mbox{NVIDIA}^{\text{\textregistered}}$ GP-GPU with compute capability
$\geq2.0$ . Double precision floating point operations are supported.
Therefore, we can adopt the terminology of NVIDIA CUDA. The design
guidelines that we are going to introduce can also be implemented
on other GP-GPUs and more generally on any Single Instruction Multiple
Data (SIMD) architectures, e.g. through OpenCL\cite{Khronos}\\
The present parallel implementation of the SC-IVR algorithm uses two
kernels, which are the \textit{Evolution Kernel}\texttt{ (KernelEvolution}
in the code and in Fig. \ref{fig:structureCUDA}) and the Spectrum
Kernel (\texttt{KernelSpectrum} in the code and in Fig. \ref{fig:structureCUDA}).
In the Evolution Kernel the cycle over the trajectories (see Fig.
\ref{fig:structureCPP}) is distributed over a number of $n_{traj}$
threads. The $j$-th thread evolves a given initial condition $\left(\mathbf{p}_{j}(0),\mathbf{q}_{j}(0)\right)$
from time $0$ to time $T$ and works on its private copy of the working
variables used in the sequential code. Details about the memory usage
will be presented below. In order to avoid instruction branching,
which is highly detrimental in the SIMD setup, all the trajectories
are evolved up to time $T$. The trajectory status is monitored by
means of a flag variable which is initialized to \textit{good} and
switched to \textit{bad} as soon as the determinant of the monodromy
matrix associated to the trajectory deviates from the allowed tolerance\@.
Information about the spectrum contributed by each trajectory during
its evolution are stored in its private copy of the buffer array.
We organize these private copies into the $n_{traj}\times k$ buffer
matrix (\texttt{Traj-WaveLength} in Fig. \ref{fig:structureCUDA}).
\\
When the Evolution kernel terminates, the Spectrum kernel is launched
with $k$ threads. At the end of the time evolution, the $j$-th thread
computes the weighted sum of the elements of the $j$-th column of
the buffer matrix. The results of \textit{bad} trajectories are not
considered by setting their weight to zero. The structure of the CUDA
code just described is shown in Fig. \ref{fig:structureCUDA}. The
main advantage of using two separate kernels is that we are able to
take into account the different dimensions of the problem, that is
the number $n_{traj}$ of MC trajectories and the number $k$ of sampled
energies. Another advantage is that the threads work always on separate
memory locations, making therefore unnecessary the use of atomic operations
or any other kind of thread synchronization mechanism that otherwise
would slash the performance of the parallel code.

A central task in the development of CUDA coding concerns the optimization
of the use of different types of GPU memories. As a matter of fact,
memory bandwidth can be the real bottleneck in a GPGPU computation.
As mentioned above, in order to reproduce the independence on the
MC trajectory in the code, each thread works on a private copy of
variables. On one hand, this could be easily accomplished by reserving
to each thread a portion of consecutive Global Memory words large
enough to contain all its working variables. On the other hand, this
naive approach would lead to highly misaligned memory accesses and
to a large amount of unnecessary and costly memory traffic. In order
to allow for coalesced read/write operations in the global memory,
we store the $n_{traj}$ copies of the same variable in contiguous
positions. For instance, the momenta of the different MC trajectories
are stored as $\left(p_{1}^{1},p_{2}^{1},...,p_{n_{traj}}^{1},...,p_{1}^{F},p_{2}^{F},...,p_{n_{traj}}^{F}\right)$, where $F$ is the number of degrees of freedom (the normal modes)
of the molecule. In this way neighbor threads will issue read/write
memory requests to neighbor memory locations that can be \textquotedbl{}simultaneously\textquotedbl{}
served.

We recall that atom masses, equilibrium positions, the conversion
matrix and other potential-structure matrices are constant and trajectory-independent.
Thus, we store these parameters in the Constant Memory. In this way,
when all the threads in an half-warp issue a ``read'' of the same
constant memory address, i.e. for the same parameter, a single ``read''
request is generated and the result is broadcasted to all the requiring
threads. Moreover, since constant memory is cached, all the subsequent
requests of the same parameter by other threads will not generate
memory traffic.

Finally, we discuss the matter of the L1-Cache/shared memory usage.
This 64kB memory is located close (on chip) to the processing units
(CUDA cores) and provides the lowest latency times. By default, 16kB
of this memory are used as L1-Cache memory, which is automatically
managed by the device, whereas the remaining 48kB can be used either
to share information between the threads in a block or as a \textquotedbl{}programmable
cache\textquotedbl{}. Since there is no flow of information between
threads, we use the shared memory as \textquotedbl{}programmable\textquotedbl{}
cache. Due to the limited size of this memory, we employ it to store
only the position vectors of the trajectories in a block and some
intensively used parameters. Fig. (\ref{fig:figureMemoryCUDA}) shows
where the main data structures employed by the code are allocated.

We conclude this Section with a remark about the block-versus-thread
structure. The threads that are used to generate the MC trajectories
or to compute the components of the spectrum are evenly distributed
among the blocks. The number of blocks, therefore, determines the
threads/block ratio. Taking into account the dimension of the scheduling
unit (warp), we constrain the number of threads-per-block to be an
integer multiple of 32. Subsequently, we choose the configuration
that provides the best performance, i.e. the shortest computing time.
This procedure allows us to slash most of memory latency times and
guarantees a sufficient number of Registers for each thread as well.

\section{Results and discussion\label{sec:Results-and-discussion}}

Initially, debugging calculations are performed with GPU $\mbox{NVIDIA}^{\text{\textregistered}}\mbox{Tesla}^{\mbox{TM}}\mbox{C2075}$
and CPU Intel core i5-3550 (6M Cache, 3.3GHz)  processors. Then, performance
calculations are done employing the GPU $\mbox{NVIDIA}^{\text{\textregistered}}\mbox{Kepler}^{\mbox{\text{\textregistered}}}\mbox{K20}$
and CPU Intel Xeon E5-2687W (20M Cache, 3.10 GHz) at the Eurora cluster of the Italian
supercomputer center CINECA. 
%
%
\\
In order to avoid any accidental over-estimation of the GPU code performance,
we stress that the CPU code uses a single thread and does not make
any use of SIMD instruction sets, such as Intel SSE. This means that
the CPU code is not designed to fully exploit the computational power
of multi-core or SSE-enabled processors. Considered the parallel nature of the described MC algorithm, a multi-thread version of the code will require in the best case  $1/k$ of the single-thread CPU time, where $k$ is the number of available cores.\\
As for GPUs, we use the same code on both Tesla C2075 and K20, with the exception of the block vs thread configuration that is set
to maximize the performance on each device. The new
functionalities introduced by the Kepler architecture (such as dynamic
parallelism and the 48K Read-Only Data Cache) are not exploited.

In order to test the performance of the CUDA SC-IVR code described
in the previous section, we look at four molecules with an increasing
number of degrees of freedom. It is important to study the time scaling
not only for increasing number of trajectories, but also for increasing
complexity of the molecular system. The chosen molecules are $\mbox{H}_{2}$,
$\mbox{H}_{2}\mbox{O}$, $\mbox{H}_{2}\mbox{CO}$, and $\mbox{C}\mbox{H}_{2}\mbox{D}_{2}$
and the number of their vibrational degrees of freedom is respectively
1, 3, 6, and 9. So, one should keep in mind that the vibrational mode
number grows at the fast pace of three times the number of atoms.
Another aspect to take into account for a proper time scaling evaluation
is represented by the potential energy subroutine adopted. For the
$\mbox{H}_{2}$ molecule, a simple Morse oscillator is employed, while
for the other molecules we use analytical potential energy surfaces
fitted to \emph{ab initio }quantum electronic energies.\cite{PES_H2O,PES_H2CO,PES_CH2D2} 

\begin{figure}
\begin{centering}
\includegraphics[width=0.5\paperwidth]{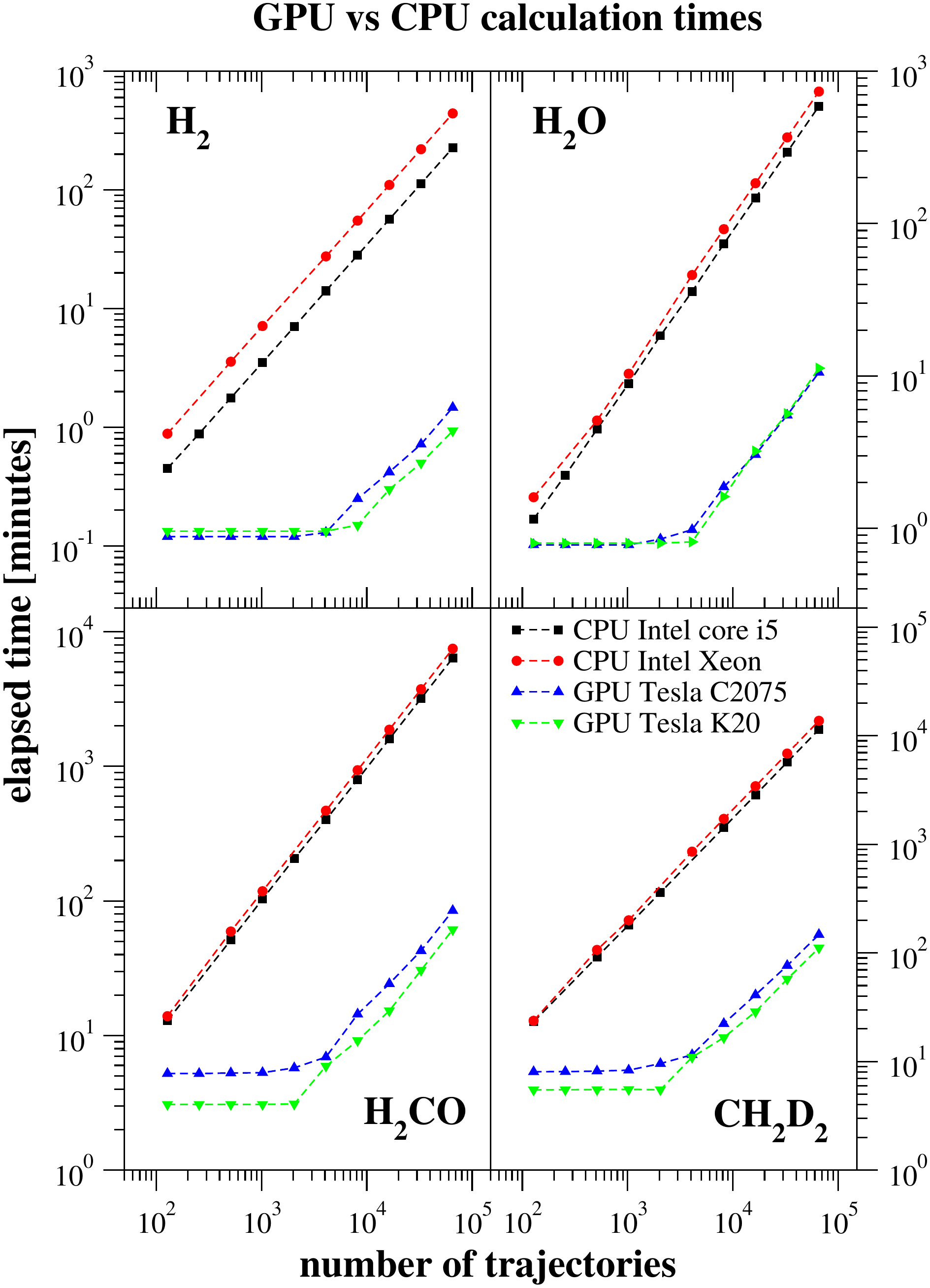}
\par\end{centering}

\caption{\label{fig:scaling_trajectories}Elapsed computational time for CPU-SCIVR
and GPU-SCIVR calculations. For a small number of trajectories (<2048), GPU times are roughly constant: the GPUs computational capabilities are not fully exploited. For a large number of trajectories (>>4096) GPU times scale linearly with respect to the number of trajectories. CPU times grow linearly with the number of trajectory on the whole range $[128, 65536]$.}
\end{figure}
We are aware that about one thousand trajectories per vibrational
degree of freedom \cite{Alex} are necessary in order to reach convergence
in the Monte Carlo integration of Eq. (\ref{eq:sep_approx}). However,
we report calculations performed up to 65536 trajectories. This allows for a study of computing capability saturation of the two GPUs under consideration (see below) as well as a better description of  the different computational simulation time trends of CPUs and GPUs.
Fig. \ref{fig:scaling_trajectories}
shows the computational time at different numbers of classical
trajectories and for different molecules.  Semiclassical CPU calculations
show a linear scaling up to the maximum number of 65536 trajectories
tested. Instead, the computational time of the SC-IVR GPU CUDA code
described above is roughly constant up to $n_{traj}=2048$ for K20 and $n_{traj}=4096$ for C2075, independently of the molecule
under investigation. While the serial operation modality enforced
by CPU architecture is clearly at the origin of the linear scaling,
the GPU behaviour is a more sophisticated one. As a matter of fact,
the execution time for a number of trajectories smaller than the indicated
thresholds (2048/4096) is very close to the time required by the GPU
to complete the evolution of a single trajectory. By accurately profiling
the execution of the code, we find that this behaviour is largely
due to the high memory traffic generated by the code, since a single
trajectory requires the manipulation of a large number of data. Thanks
to an accurate memory mapping of the information needed by the code
(see below), we are able to minimize the on-chip/off-chip data transfer.
We find that the latency-time (i.e. the amount of time required for
data to become available to a thread) is playing a central role. However,
when the number of threads is larger than the number of Streaming
Multiprocessors (the computational units in CUDA), part of the latencies
is hidden by the thread scheduler. Instead, when a thread is inactive
while waiting for data to arrive, another one, which is ready for
execution, is run. The execution time ceases to remain constant as
soon as the number of threads becomes larger than the time needed
to hide the latencies. This occurs when the computational power of
the GPU is saturated. Interestingly enough, the \textquotedbl{}more
powerful\textquotedbl{} K20 gets saturated sooner (2048 threads) than
the \textquotedbl{}old\textquotedbl{} C2075. We will address this
issue later in this section. 

For every molecular system, once the number of trajectories is large
enough for the Monte Carlo integral to converge, the resulting power
spectrum is compared to the one reported by Kaledin and Miller.\cite{Alex}
We find our eigenvalues to be in agreement within 0.1\%. This negligible
discrepancy is due to the slightly different number of trajectories
used in the GPU calculations.
%
\begin{figure}
\begin{centering}
\includegraphics[scale=0.4]{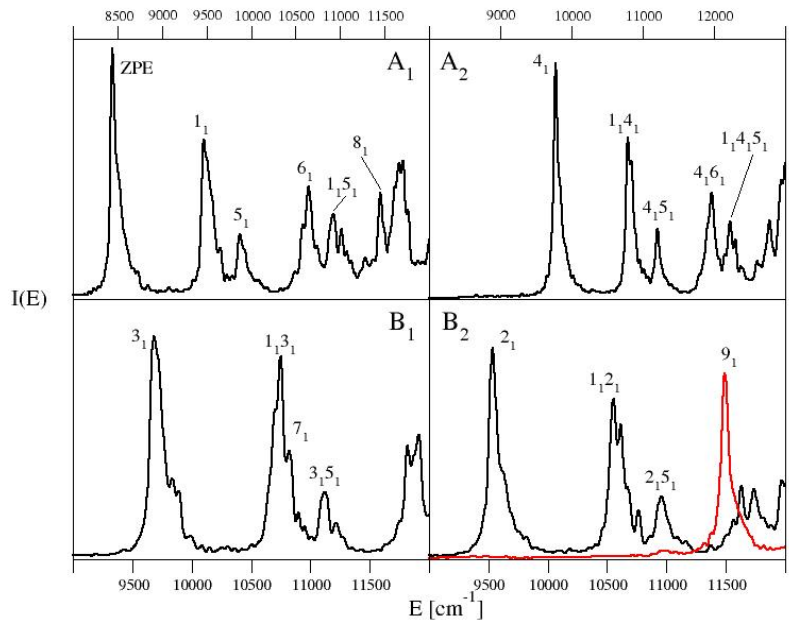}
\par\end{centering}

\caption{\label{fig:CH2D2}
The deuterated methane $\mbox{CH}_{2}\mbox{D}_{2}$ power spectrum using the K20 GPU.
The MC integration was converged with 8192 trajectories and the spectrum
has been projected onto the four irreducible representations for a
peak attribution. Red and black lines are different coherent state
combinations for the same irreducible representation.}
\end{figure}
 As an example, we report in Fig.
  \ref{fig:CH2D2}
the power spectrum of di-deuterated methane. We stress once more that our main goal is to test accuracy
and efficiency of the GPU implemented SC-IVR code and not just the
determination of the spectrum, a problem which has already been solved.
In Fig.
\ref{fig:CH2D2}, the power spectrum  of the deuterated methane $\mbox{CH}_{2}\mbox{D}_{2}$ is
projected onto the irreducible representations of the relevant molecular
point group. This procedure helps the reader, assists the authors
to assign peaks more easily and permits a stricter comparison with
previous calculations on the same systems.\cite{Alex,Ceotto_cursofdimensionality_11}
%

\begin{figure}
\begin{centering}
\includegraphics[scale=0.4]{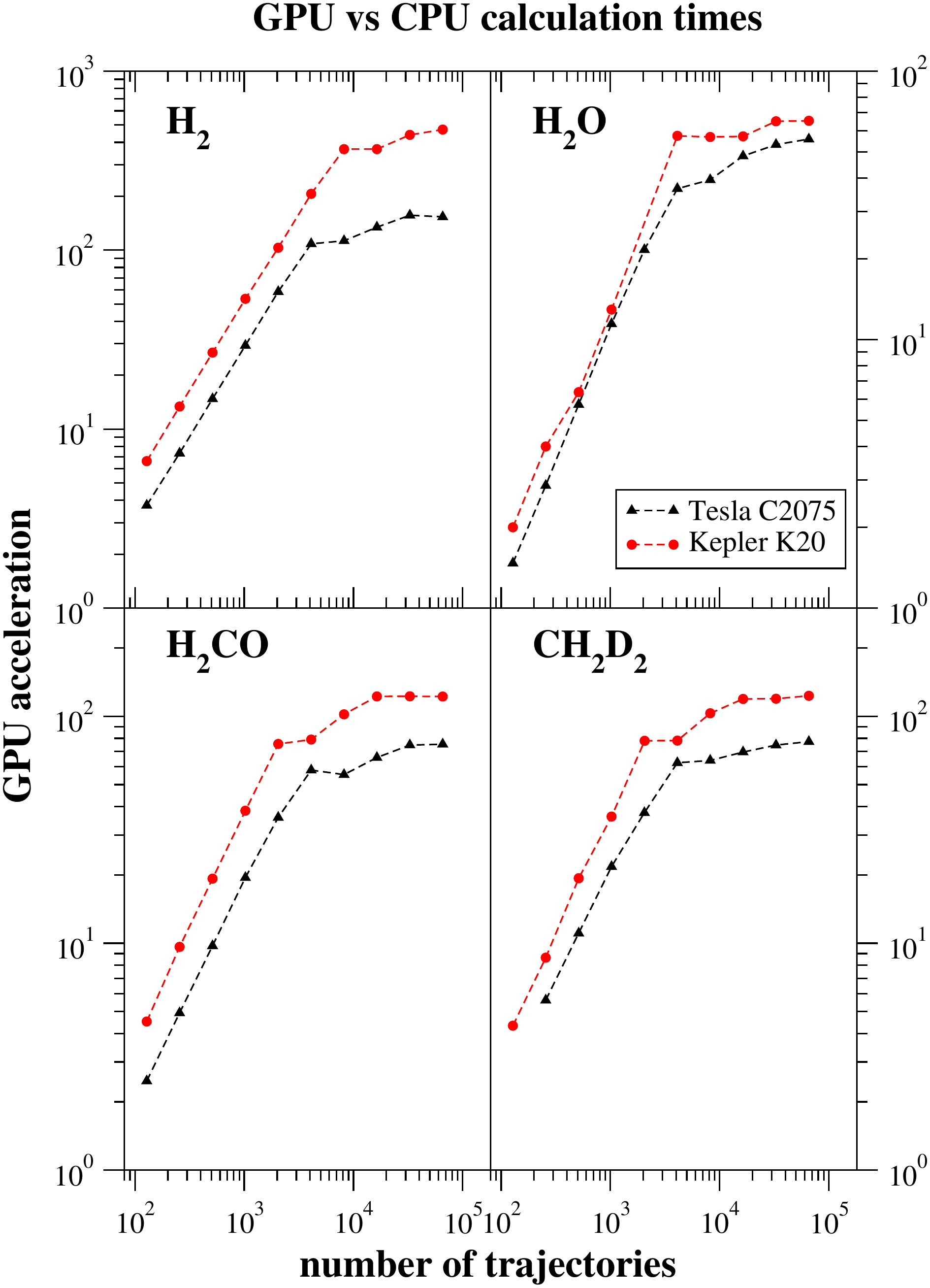}
\par\end{centering}

\caption{\label{fig:C2075vsK20}NVIDIA graphics units C2075 and K20 performances
respectively versus the Intel Core i5 and the Intel Xeon E5-2687W
for the calculation of the power spectra.}
\end{figure}

After verifying that indeed the GPU implemented code preserves the
same accuracy of the CPU one, we turn to the computational performance
difference between the two NVIDIA graphics units, i.e. C2075 and
K20.
Clearly, for each molecule tested,
one would expect a better performance of the more recent K20 with
respect to C2075, as reported in Fig. \ref{fig:C2075vsK20}. 
 The acceleration amount shown in Fig. \ref{fig:C2075vsK20} increases
with the number of trajectories. The upper left panel of the Figure
reports the speed-up for the Morse oscillator power spectrum calculation.
The acceleration is comparable in magnitude for the water molecule
presented on the upper right panel and the two graphics units performances
are quite similar. Instead, for the complex systems reported on the
lower panels, the acceleration of the K20 graphic card is larger,
as pointed out by the log-scale.

The trends of computational time (see Fig. \ref{fig:scaling_trajectories})
and acceleration beyond $n_{traj}=2048$ for K20 and $n_{traj}=4096$
for C2075 deserve some further discussion. As mentioned above,
for a given number $n_{traj}$ of MC trajectories, the number of threads
per block configuration is always chosen to maximize the performance.
This number is usually kept high enough to make it possible for the
warp scheduler to hide memory access latency. We find out that for
$256\leq n_{traj}\leq8192$ the best results are obtained with 128
threads in each block, independently of the device we run the code
on. The real occupancy, i.e. the number of warps (execution units
of 32 threads) running concurrently on a multiprocessor, however,
is determined by the register needed by each thread. This is a key
issue when codes are using a high number of variables, like in the
present case. The dimension of the SMX register file is twice the
size of the register memory for C2075 (256kB vs. 128kB), so the
occupancy of K20 can be higher than that of C2075. On one
side, this contributes to speed-up the calculations as shown by the
K20 device performances. On the other side, the size of the L1 cache
memory we use (48kB) is the same on both devices. This means that the same
amount of L1 cache is shared among more really concurrent threads
on K20 than on C2075. This results in a higher on-chip/off-chip
memory traffic, and it is likely a reason for the earlier reduction
of the GPU acceleration growth rate of Kepler K20 with respect
to Tesla C2075.
\begin{table}[h]
\caption{\label{tab:Performances}Performance of each computing device.$^{a)}$}

\begin{centering}
\begin{tabular}{|ccccccc|}
\hline 
Molecule & i5-3550 & C2075 & ratio CPU/GPU & E5-2687W & K20 & ratio CPU/GPU\tabularnewline
\hline 
\hline 
$\mbox{H}_{2}$ & 28.20$^{b)}$ & 0.25 & 113 & 55.03 & 0.15 & 367\tabularnewline
\hline 
$\mbox{H}_{2}\mbox{O}$ & 73.63 & 1.87 & 39 & 91.60 & 1.62 & 57\tabularnewline
\hline 
$\mbox{H}_{2}\mbox{CO}$ & 798.07 & 14.42 & 55 & 936.73 & 9.18 & 102\tabularnewline
\hline 
$\mbox{CH}_{2}\mbox{D}_{2}$ & 1428.35  & 22.35 & 64 & 1715.73  & 16.63  & 103\tabularnewline
\hline 
\end{tabular}
\par\end{centering}
$a)$ The number of trajectories is 65536. $b)$ The computational time is measured in minutes.
\vspace{1cm}
\end{table}
 Table \ref{tab:Performances} reports the computational time for
the fully converged TA-SC-IVR spectra calculations for the devices
employed. This Table shows that the K20 computational time is always
smaller than that of the older C2075 and of any other CPU device.

\begin{figure}
\begin{centering}
\includegraphics[width=0.6\paperwidth]{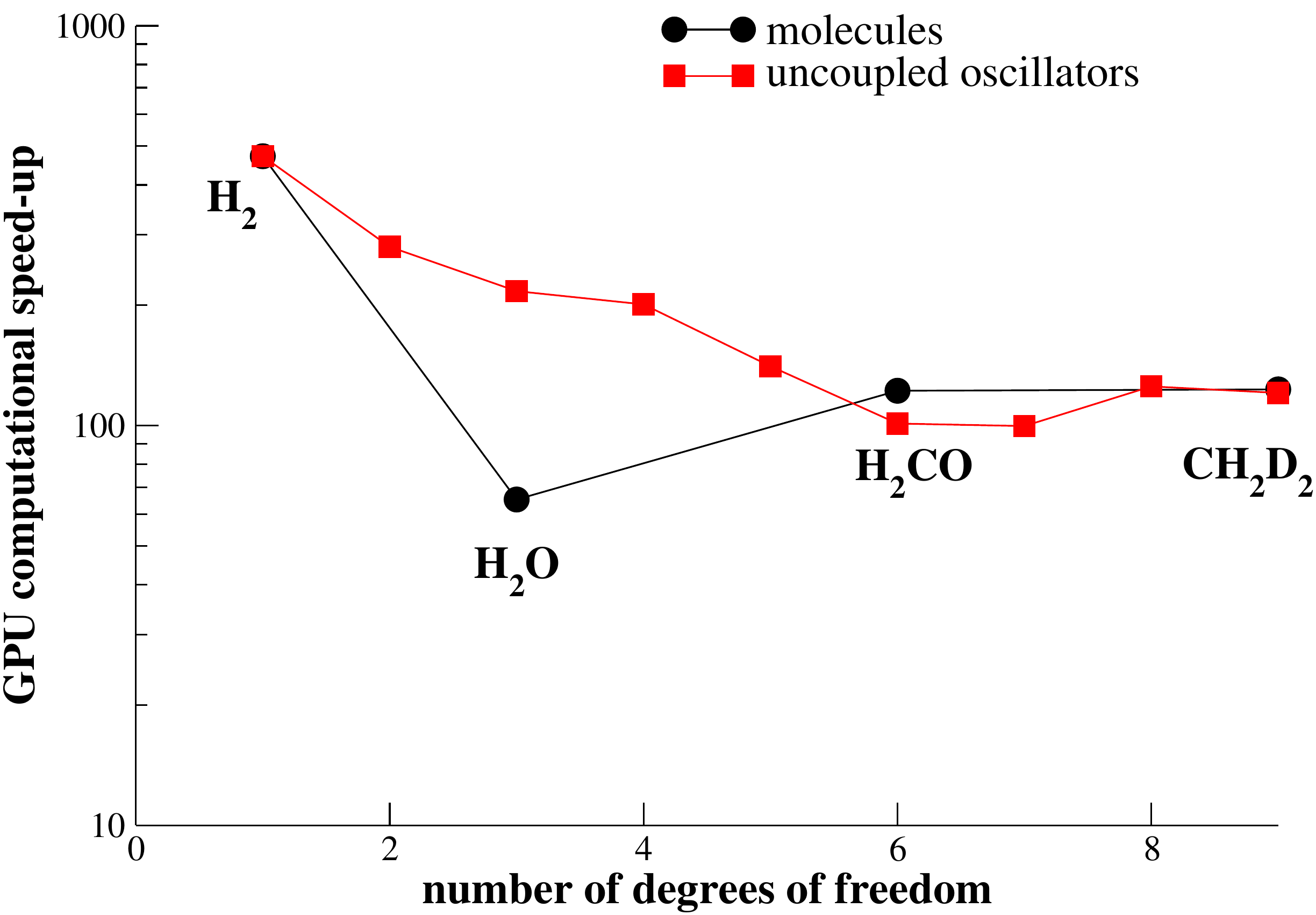}
\par\end{centering}

\caption{\label{fig:speedup}Power spectrum GPU computational speed-up for
$\mbox{H}_{2}$, $\mbox{H}_{2}\mbox{O}$, $\mbox{H}_{2}\mbox{CO}$,
and $\mbox{C}\mbox{H}_{2}\mbox{D}_{2}$ (filled black circles) compared
to uncoupled Morse oscillators (filled red squares) with the same
number of degrees of freedom and for a 65536 trajectory Monte Carlo
integration.}
\end{figure}

If we consider the computational time for the better performing K20
GPU and look at the time scaling for all the molecules under examination,
we obtain the plot reported by the black filled circles in Fig. \ref{fig:speedup}.

Opposite to what one would expect, the ratio of the CPU computational
time over the GPU one is not monotonically decreasing with the number
of vibrational degrees of freedom for the molecule calculations. To
find the source of such an irregular behavior, we treat a set of uncoupled
Morse oscillators. We calculate the ratio between the CPU and GPU
computational time for an increasing number of oscillators while keeping
the same number of trajectories used for the molecules considered.
The red filled squares in Fig. \ref{fig:speedup} report these values.
In this case, the GPU acceleration contribution is slightly decreasing
with the number of degrees of freedom. In the case of a single oscillator,
the GPU speed-up factor is exactly the same found for the hydrogen
molecule because the $\mbox{H}_{2}$ potential is a Morse potential.
Also the speed-up for the molecules with six and nine degrees of freedom
is similar to that for the corresponding oscillators. Conversely,
the GPU acceleration for the water molecule strongly deviates from
its Morse oscillator reference. We ascribe this speed-up discrepancy
to the potential subroutine. This subroutine is called several times,
i.e. at each time step and for each trajectory. For a typical 65536
trajectory simulation with a fourth order symplectic algorithm iterated
for 4000 time steps, the potential subroutine is called about $2.9\times10^{11}$
times and we estimate it to take, for all the molecules except $\mbox{H}_{2}$,
approximately 70\% of the overall running time. Different analytical
expressions for the fitting potential surfaces lead to different performances
after GPU implementation. For instance, an additional square root
calculation can significantly change the computational time considered
the number of times the potential subroutine is called. Thus, Fig.
\ref{fig:speedup} eloquently shows how important it is to write the
potential energy surface in an analytical form as simple as possible.
We actually think that this consideration is valid beyond the employment
of the GPU hardware. These limitations related to the fitted analytical
potential are not present in a direct ``on-the-fly'' semiclassical
dynamics simulation. However, in this last case, the bottleneck is
represented by the cost of \emph{ab initio }electronic energy calculations,
especially when high level electronic theory and large basis sets
are employed. A viable and convenient future perspective would be
to combine the present SC-IVR Monte Carlo GPU parallelization with
the available GPU \emph{ab initio} codes,\cite{Martinez_2008} to
investigate if and to which extent GPUs slash direct dynamics times
and allow accurate calculations for sizeable molecules.

Finally, we discuss the power consumption convenience of using GPU devices for semiclassical calculations. Thanks to the support of the Italian Supercomputing Center CINECA, we have been able to measure the amount of energy dissipated by each job. As an example, we focus on 65536 trajectory runs for deuterated methane, which is the largest molecule considered in this work. We found that the power dissipated by the K20 GPU computation is 0.33 kWh, whereas for the single-thread CPU computation on Xeon is 24.50 kWh. Even assuming that an eight concurrent threads simulation is consuming $24.50/8$ kWh, the GPU run is still ten times more convenient, in terms of power consumption, than the CPU one.

\section{Conclusions}
This paper describes the implementation of the SC-IVR algorithm for CUDA GPUs. Through a careful usage of the memory hierarchy, it is possible to use a GPU as if it were a ``cluster'' of CPUs, each working on an independent memory space. We find a significant speed-up with respect to CPU simulations. Taking a  multi-thread simulation over eight cores, the GPU speed-ups is lowered to about 12 for most of the molecules here considered. Interestingly enough, the performance delivered by the GPU is strongly dependent on the kind of operations required by the potential energy surface subroutine. We bench-marked the code on molecules up to nine degrees of freedom. Our future work will be mainly focused on the development of new implementations able to offer a viable alternative route to the use of multiple parallel GPUs in applications where a large number of trajectories is necessary.


\section*{Acknowledgement}

The authors thank NVIDIA\textquoteright{}s Academic Research Team
for the grant MuPP@UniMi and for providing the Tesla C2075 graphic
card under the Hardware Donation Program. We acknowledge the CINECA
and the Regione Lombardia award under the LISA initiative (grant MATGREEN),
for the availability of high performance computing resources and support.
Prof. A. Lagana, Prof. E. Pollak, and Prof. A. Aspuru-Guzik are warmly
thanked for useful advices and fruitful discussions. X. Andrade and
S. Blau are thanked for revising the manuscript.

\end{document}